\documentclass{svproc}
\usepackage{url}
\usepackage{amsmath}
\usepackage[linesnumbered,ruled,vlined]{algorithm2e}
\usepackage{algpseudocode}
\usepackage{graphicx}
\usepackage{xspace}
\usepackage{amssymb}

\begin{document}

\mainmatter              
\title{Benchmarking Metaheuristic-Integrated QAOA against Quantum Annealing}
\titlerunning{Metaheuristic-Integrated QAOA v. QA}  
%
\author{Arul Rhik Mazumder\inst{1} \and Anuvab Sen \inst{2} \and Udayon Sen\inst{2}}
\authorrunning{Arul Rhik Mazumder et al.} 
%
\tocauthor{Arul Rhik Mazumder and Anuvab Sen and Udayon Sen}
\institute{Carnegie Mellon University, Pittsburgh PA 15289, USA,\\
\email{arulm@andrew.cmu.edu}
\and
Indian Institute of Engineering Science and Technology, Shibpur  711103, India,\\
\email{sen.anuvab@gmail.com}, \email{udayon.sen@gmail.com}}
\maketitle              

\begin{abstract}
The Quantum Approximate Optimization Algorithm (QAOA) is one of the most promising Noisy Intermediate Quantum (NISQ) Algorithms in solving combinatorial optimizations and displays potential over classical heuristic techniques. Unfortunately, QAOA's performance depends on the choice of parameters and standard optimizers often fail to identify key parameters due to the complexity and mystery of these optimization functions. In this paper, we benchmark QAOA circuits modified with metaheuristic optimizers against classical and quantum heuristics to identify QAOA parameters. The experimental results reveal insights into the strengths and limitations of both Quantum Annealing and metaheuristic-integrated QAOA across different problem domains. The findings suggest that the hybrid approach can leverage classical optimization strategies to enhance the solution quality and convergence speed of QAOA, particularly for problems with rugged landscapes and limited quantum resources. Furthermore, the study provides guidelines for selecting the most appropriate approach based on the specific characteristics of the optimization problem at hand.
\end{abstract}
\section{Introduction}
Quantum computing is a promising emerging technology that utilizes the properties of quantum mechanics such as superposition, entanglement, and interference to provide speedups and solutions to certain classical intractable problems across various sectors. One such field conducive to quantum speedup is combinatorial optimization. Combinatorial Optimization is the process of finding optimal solutions to an objective function over a large, but discrete search space.  Informally, this may be thought of as making optimal choices from a set of possibilities, where each choice can have a significant impact on the overall outcome. It has a diverse set of applications across countless fields such as logistics, supply chain optimization, bio-informatics, game theory, and machine learning, and is actively researched in Operations Research, Mathematics, and Computer Science. In particular, the research focuses on solving commonly used Quadratic Unconstrained Binary Optimization (QUBO) problems \cite{DBLP:journals/corr/abs-1811-11538}. These models are primarily used in quantum computing due to their simplicity and equivalency to Ising models. Ising models are physics models used to represent the behavior of interacting spins in physical systems. Because of its connection to physics, it is very easy to map the logical qubits in an Ising model to the physical qubits of a quantum processing unit (QPU) during the embedding stage of quantum computing. QUBO models maintain the binary properties of computer science while also retaining a simple topological mapping to physical qubits. There have been two main quantum algorithms to solve QUBO models, Quantum Approximate Optimization Algorithm (QAOA) \cite{farhi2014quantum}\cite{sturm2023theory} and Quantum Annealing (QA) \cite{de_Falco_2011}. These two algorithms are unique as they have each been implemented for different quantum computing models: QAOA for gate-based computing and QA for adiabatic quantum computers. In past works, QA outperforms stand QAOA in terms of solution quality \cite{Pelofske_2023}. There have been no studies documenting the comparison in time to attain these higher solution qualities. Recently there has been extensive research into the application of metaheuristic algorithms as classical optimizers for QAOA circuits. Metaheuristic algorithms are general-purpose optimization algorithms that use high-level strategies to explore the search space to identify optimal or near-optimal solutions. These algorithms are often inspired by natural processes or phenomena with some of the most famous metaheuristic algorithms being Differential Evolution \cite{storn1997differential}, Genetic Algorithm \cite{Holland1984}, and Particle Swarm Optimization \cite{488968}. Past research works have often found that metaheuristic-integrated QAOA circuits often return higher-quality solutions than vanilla QAOA circuits \cite{9555505}, \cite{faílde2023using}, \cite{miranda2021synthesis}, \cite{sünkel2023ga4qco} however no research compares the metaheuristic-integrated QAOA circuits to Quantum Annealing.
In this research paper, we focus on benchmarking these novel optimized metaheuristic-optimized QAOA circuits against quantum annealing in solving QUBOs using the metrics of classical and quantum execution time and solution approximation ratio. Although there is extensive research comparing gradient-based and gradient-independent optimizers, there are few papers investigating metaheuristic optimizers. Metaheuristics have the advantage over standard optimizers in the ability to handle non-standard functions, and global search capabilities, and trading off breadth vs. depth exploration. Thus metaheuristic-optimizers will identify better parameters for QAOA circuits and the resulting circuits will have higher approximation ratios. The metaheuristics algorithms used in this paper are the Genetic Algorithm (GA), Differential Evolution (DE), Particle Swarm Optimization (PSO), and Ant Colony Swarm Optimization (ACSO)\cite{dorigo2006ant}, and no previous literature investigates such a large and diverse set of metaheuristic optimizers. The primary goal of the research work is to evaluate and compare the effectiveness and efficiency of these two quantum approaches for scalable and practical QUBO problem-solving tasks. Overall, this research strives to provide valuable insights into the performance and long-term potential of QAOA and quantum annealing techniques in solving QUBOs, providing intuition into the best quantum algorithms and best quantum computing platforms to solve the difficult optimization problems of the future.

\section{Background Work}
\subsection{Number Partitioning Problem}
The paper benchmarks each algorithm's performance in solving the Number Partitioning Problem (NPP) \cite{mertens2003easiest}. The Number Partitioning problem is defined as follows: Given a set $S$ of positive $n$ integers $\{a_1, a_2, a_3, ... , a_n\}$, partition the set into two disjoint subsets $A$ and $S/A$ such that:
\begin{equation}
|\sum_{a_i \in A}a_{i} - \sum_{a_j \in S/A}a_{j}| = d
\end{equation}
where $d$ is minimized. 

If $d=0$, the set is perfectly partitioned. This optimization problem is NP-Hard and grows exponentially with the size of the input, but there exists a pseudo-polynomial time dynamic programming algorithm which means the time grows polynomially with the largest input value. 

The Number Partitioning Problem was chosen for this research for two core reasons. First NPP is one of Garey and Johnson's six core NP-hard problems and NPP serves as the base for other challenging NP-hard problems like Knapsack and Bin Packing. NPP itself has applications from multiprocessor scheduling to public key cryptography. Thus it is one of the most fundamental and practical NP-Hard problems. Secondly, the poor quality of modern heuristics. Unlike other NP-Hard problems like the Traveling Salesman Problem \cite{LAPORTE1992231}, the best polynomial time heuristic algorithms yield discrepancies of $O(N^{-\alpha \log N})$ which is significantly worse than the best case. Thus due to the poor state of modern classical heuristics, quantum algorithms even in the Noisy Intermediate Scale Quantum Era (NISQ) may show promising improvements.

\subsection{Quantum Annealing}
Quantum Annealing (QA) is a pioneering optimization technique that applies quantum mechanics principles to solve complex optimization problems. It uses quantum tunneling and entanglement, fundamental principles in quantum mechanics, to comb through solution spaces more efficiently than classic optimization techniques. It is based on the annealing process, a phenomenon where a quantum system's energy is gradually transitioned from a higher-energy state to a low-energy state. QA manipulates quantum bits (aka qubits), the fundamental units of quantum computation. The qubits are evolved using Hamiltonian operators, which guide the system to the ground state. The optimal solution is encoded in the final quantum state, representing the optimal solutions' probability distribution. The pseudocode for the Quantum Annealing (QA) algorithm is provided in Algorithm $1$.

\RestyleAlgo{ruled}
\begin{algorithm}
\caption{Quantum Annealing}\label{alg:cap}
\SetAlgoLined
\KwData{Problem Hamiltonian $H$, Annealing Schedule $\tau$, Number of Annealing Steps $T$}
\KwResult{Solution State $|\psi_{\text{final}}\rangle$}
Initialize a quantum system in an initial state $|\psi\rangle$\;
\For{$t = 1$ \KwTo $T$}{
    Calculate the annealing parameter $s = t / T$\;
    
    Generate the time-dependent Hamiltonian $H_t = (1 - s) \cdot H_0 + s \cdot H_P$, where $H_0$ is the initial Hamiltonian and $H_P$ is the problem Hamiltonian\;
    
    Evolve the quantum system according to $H_t$ for a time step $\Delta t$ using a quantum gate or simulation technique\;
}
Measure the quantum state $|\psi\rangle$ to obtain a classical bit string\;
Return the classical bit string as the solution state $|\psi_{\text{final}}\rangle$\;
\end{algorithm}

\subsection{Quantum Approximate Optimization Algorithm (QAOA)}

The Quantum Approximate Optimization Algorithm (QAOA) is a novel gate-based quantum computing approach created to tackle combinatorial optimization problems more efficiently than classic optimization techniques. It is a hybrid quantum-classical that searches for approximate solutions to combinatorial optimization problems. 
It operates by evolving a quantum state with the guidance of a classical optimization algorithm. The QAOA has two steps: the evolution of the quantum operator and classical optimization. 
The pseudocode for Quantum Approximation Optimization Algorithm is given in the Algorithm $2$.

\RestyleAlgo{ruled}
\begin{algorithm}
    \caption{Quantum Approximate Optimization Algorithm}
    \SetAlgoNlRelativeSize{-1}

    \KwIn{Problem Hamiltonian $H_p$, Mixer Hamiltonian $H_m$, Number of QAOA steps $p$}
    \KwOut{Quantum state $\psi_{\text{best}}$}

    Initialize quantum circuit $U(\beta, \gamma)$ with random parameters $\beta$ and $\gamma$\;
    $\psi_{\text{best}}$ = initial state (e.g., $|000...0\rangle$)\;

    \For{$k = 1$ \KwTo $p$}{
        Apply $U(\beta_k, \gamma_k)$ to $\psi_{\text{best}}$ // Apply the parameterized quantum circuit\;
        $\psi_{\text{best}}$ = Evolve($H_p$, $H_m$, $\psi_{\text{best}}$) // Time-evolve using $H_p$ and $H_m$\;
        Measure expectation value $E_k = \langle \psi_{\text{best}} | H_p | \psi_{\text{best}} \rangle$\;

        Compute gradients of $E_k$ w.r.t. $\beta_k$ and $\gamma_k$\;
        Update $\beta_k$ and $\gamma_k$ using an optimization method (e.g., gradient descent)\;
    }
    \KwRet $\psi_{\text{best}}$\;
\end{algorithm}
The operators are problem constraints, and the variational parameters control the evolution, while the classical optimization adjusts the parameters to improve the solution quality iteratively. We employ the use of a Hamiltonian operator for QAOA.

Hamiltonian Operator basically corresponds to the total energy of the system. The Hamiltonian operator is a linear operator which applies a Hamiltonian transformation to the existing function.

For QAOA, we are defining two Hamiltonians, based on our requirements. Firstly, the $problem$ Hamiltonian $H_p$ is such that its ground state references the solution of the optimization problem. Then we have the $mixer$ Hamiltonian $H_m$, which plays a big role in guiding the quantum state towards optimal solutions while applying the optimization technique of our choice.  We create circuit layers based on these two parameters and then employ the optimization techniques to get the optimal solution quality.

\subsection{Differential Evolution}

Differential Evolution (DE) is a population-based metaheuristic that solves non-differentiable and non-linear optimization problems employing a unique mutation and selection strategy to optimize the candidate solutions over iterations. DE obtains an optimal solution by maintaining a population of possible candidate solutions and iteratively improving these solutions by applying genetic operators. DE employs three main operators: mutation, crossover, and selection. Like the Genetic Algorithm, Differential Evolution begins by randomly initializing a population and then generating new solutions using crossover and mutation operators. A mutation is a stochastic change to a candidate solution parameters. Similar to the genetic algorithm, mutations are utilized to increase diversity in the candidate solutions to prevent the algorithm from converging upon a sub-optimal local optimum. The original mutation scheme devised by Storn creates a mutant vector $v_{i}$ by randomly sampling three candidate solution vectors $x_{r_{1}}$, $x_{r_{2}}$, $x_{r_{3}}$ according to Equation $3$
\begin{equation}
    v = x_{r_{1}} + F \times (x_{r_{2}} - x_{r_{3}})
\end{equation}
In the equation above $F$ represents the scale factor controlling the magnitude of the mutation. Increasing F enhances DE’s explorative capabilities at the expense of its efficiency while decreasing F does the opposite. The crossover operator combines the mutated individual with the targeted individual to create trial solutions, while the selection operator decides whether the trial solution replaces the target individual in the population. The crossover technique allows the following generation to inherit characteristics from the mutant vector. The crossover scheme employed in this work is described in Equation $4$.
\cite{Georgioudakis_Plevris_2020}. 
\begin{equation}
u_{j, i} = 
\left\{
    \begin{array}{lr}
        v_{j, i}, & \text{if } p_{rand}(0, 1) \leq CR \\
        x_{j, i} & \text{else} 
    \end{array}
\right\}
\end{equation}
The selection factor determines which solution to pass by measuring the quality of each candidate vector using objective function $f$. 
The selection operator devised for the simulation can be described in Equation $5$ \cite{Wang_2003}.\

\begin{equation}
x^{g+1}_{i} = 
\left\{
    \begin{array}{lr}
        u^{g}_{i}, & \text{if } f(u^{g}_{i}) \geq f(x^{g}_{i}) \\
        x^{g}_{i} & \text{else} 
    \end{array}
\right\}
\end{equation}

The pseudocode for the Differential Evolution is provided in Algorithm $3$.

\RestyleAlgo{ruled}
\begin{algorithm}
    \caption{Differential Evolution}
    \label{algo:de}
    \SetAlgoNlRelativeSize{-1}

    \KwIn{Population Size $N$, Dimension $D$, Scaling Factor $F \in (0, 2)$, Crossover Probability $CR \in [0, 1]$, Termination Criterion}
    \KwOut{Best Individual}
    Initialize the population $P$ with $N$ random individuals in the search space\;
    \While{Termination Criterion is not met}{
        \ForEach{individual $i$ in $P$}{
            Select three distinct random individuals $a$, $b$, and $c$ from $P$\;
            Generate a trial vector $v$ by combining the components of $a$, $b$, and $c$ using the DE mutation strategy\;
            \ForEach{dimension $j$ in $D$}{
                Generate a random number $r \in [0, 1]$\;
                \If{$r < CR$ or $j$ is a random dimension}{
                    $v[j] = v[j]$\;
                }
                \Else{
                    $v[j] = i[j]$\;
                }
            }
            Evaluate the fitness $f(v)$\;
            \If{$f(v) < f(i)$}{
                Replace individual $i$ with trial vector $v$\;
            }
        }
    }
    \KwRet Best individual in the final population\;
\end{algorithm}
Differential evolution (DE) thus operates by iteratively updating the candidate solutions and evolving the population over multiple generations iteratively. The process is repeated for a specific number of generations, until a termination criterion is satisfied, or a desired level of convergence is achieved eventually.

\subsection{Genetic Algorithm}

Genetic Algorithm (GE) is a bio-inspired metaheuristic algorithm based on the concept of natural selection, which is the cause of biological evolution as espoused by Darwin's theory of evolution. The way evolution rewards successful individuals in a population, the GE generates optimal solutions in a constrained environment.  
GE simulates the process of evolution within a population of potential solutions. Each solution was referred to as individuals who go through operations like selection, crossover and mutation, reproduction, and inheritance similar to organisms in the natural world. As multiple generations pass, the GA develops better solutions, ultimately converging toward the optimal outcome. There are four general steps in the process of Genetic Algorithm. 

\paragraph{Initialization:}

A population with size $N$ where $N \in (5,10)$ is initialized. In our case, the population size was chosen because this was the minimum size which showed an optimum convergence. Each individual in the population is a candidate solution and has a unique set of genes known as a chromosome, of length $L$ which is represented as a binary string $\{0, 1\}^k$ where $k$ is the length of the chromosome \cite{Katoch_Chauhan_Kumar_2020}. Each gene is expressed as $0$ or $1$ and represents a variable characteristic that can be manipulated to adjust the solution's optimality. The genes come together to form the chromosome. During the initialization, individuals are created, and their genes are set randomly to allow the algorithm to search through the maximum range of possible solutions. 

\paragraph{Selection:}

Once the population initialization is complete, the candidate solutions are evaluated with a fitness function $f$. The fitness function determines the solution's quality based on the evaluated chromosomes. Candidates with higher fitness are more likely to be chosen in this step to pass on their genomic characteristics to their offspring. In contrast, the candidates with poor fitness are more likely to fail, which leaves only higher-quality solutions for each iteration. Because of its nature, the selection operator is also called the reproduction operator, and the rate of convergence of GA towards an optimal solution depends on the selection pressure determined by the fitness function \cite{Sünkel_Martyniuk_Mattern_Jung_Paschke_2023}.

For our case, we are implementing the Roulette Wheel Selection method. The idea of the method is to emulate a roulette wheel, where the probability of being selected is based on the fitness of the candidate solution. 

Mathematically, the Roulette Wheel Selection expression is shown in Equation 1. 

\begin{equation} 
\frac{\text{fitness}(i)}{\sum_{j} \text{fitness}(j)}
\end{equation}

\paragraph{Crossover:}

With the candidates chosen from the selection step, two other genetic operators are used to generate offspring solutions stochastically. The first operator is the crossover operator, which is inspired by the phenomenon of chromosomal crossover occurring during sexual reproduction. Among the multiple crossover schemes, the one we have utilized is uniform crossover \cite{Syswerda_1993}, where every genomic characteristic $j$ passed to the child $c_i$ is chosen at equal probability from each parent $p_1$,$p_2$, shown in Equation $7$:

\begin{equation}
c_{j, i} = 
\left\{
    \begin{array}{lr}
        p_{1_{j, i}} & \text{if } p_{rand}(0, 1) \leq 0.5 \\
        p_{2_{j, i}} & \text{else} 
    \end{array}
\right\}
\end{equation}

\paragraph{Mutation:}

The last step of the Genetic Algorithm is critical in generating an ideal solution for optimal convergence. It emulates the biological phenomenon of mutation, a natural way of increasing diversity in a population pool with random stochastic gene-level changes, leading to better candidate solutions for convergence.
In the scope of computer science, mutations prevent the GA from converging on a local minimum solution. The summarized pseudocode for the Genetic Algorithm is displayed in the Algorithm $4$.

\begin{algorithm}
    \caption{Genetic Algorithm}
    \SetAlgoNlRelativeSize{-1}
    
    \KwIn{Population Size $N \in (5,10)$, Chromosome Length $L$, Termination Criterion}
    \KwOut{Best Individual}

    Initialize the population with $N$ random individuals\;
    \While{Termination Criterion is not met}{
        \ForEach{individual $x_i$ in population}{
            Evaluate fitness $f(x_i)$ for individual $x_i$\;
        }
        Select parents $p_1$, $p_2$ from the population for mating using a Roulette Wheel Selection scheme\;
        \ForEach{selected parent pair $(p_1, p_2)$}{
            Apply uniform crossover and mutation operations on parents $p_1$ and $p_2$ to obtain child $c_i$\;
        }
        Replace the current population with the new population\;
    }

    \KwRet Best individual in the final population\;
\end{algorithm}

Each generation cycles through the four phases until GA iterates through the maximum number of cycles or until a termination criterion is met. 

\subsection{Particle Swarm Optimization}
Particle Swarm Optimization (PSO) is a bio-inspired metaheuristic optimization algorithm based on the behavior of a fish or bird swarm in nature. It is a popular choice for solving complex optimization problems owing to its efficiency and simplicity. It aims to search and find the optimal solution in a multidimensional search space by simulating the pattern and movement of particles. Like other metaheuristics, PSO begins with the initialization of particles with velocity and position set arbitrarily. Each particle is a representation of a solution. The particle positions and velocities are updated iteratively based on a fitness function and a global best solution found by the swarm. The summarized pseudocode for the Particle Swarm Optimization Algorithm is displayed in Algorithm $5$.

\begin{algorithm}
\caption{Particle Swarm Optimization (PSO)}
\SetKwInput{Input}{Input}
\SetKwInput{Output}{Output}

\Input{Population Size $N$, Number of Iterations $MaxIter$, Problem-specific Initialization, Termination Criterion}
\Output{Best Particle}

Initialize particles' positions and velocities in the search space\;
\For{$iteration = 1$ to $MaxIter$}{
    \For{each particle $i$}{
        Evaluate fitness function $f(x_i)$ for particle $i$\;
        \If{$f(x_i)$ is better than the best fitness value of particle $i$}{
            Update personal best position: $pbest[i] = x_i$\;
            Update personal best fitness value: $pbest\_value[i] = f(x_i)$\;
        }
    }
    Update global best particle: $gbest = $ particle with the best fitness among all particles\;
    \For{each particle $i$}{
        Update particle velocity and position using equations:\;
        $velocity[i] = inertia \times velocity[i] + cognitive\_coefficient \times random() \times (pbest[i] - position[i]) + social\_coefficient \times random() \times (best - position[i])$\;
        $position[i] = position[i] + velocity[i]$\;
    }
}
\Return $gbest$\;
\end{algorithm}

Particle Swarm Optimization candidate solutions get iteratively improved using a cost function. This function takes a candidate solution as an argument in the form of a vector of real numbers and produces a real number as output. The goal is to find a solution \( a \) for which \( f(a) \leq f(b) \) for all \( b \) in the search space, where \( a \) is the global minimum. In Algorithm $5$, $b_{lo}$ \& $b_{up}$ are bounds on the range of values within which the initialization parameters will be initialized. As evident Particle Swarm Optimization Algorithm does not use the gradient of the problem being optimized, which means PSO does not require that the optimization problem be differentiable as is required by other classic optimization methods.

\subsection{Ant Colony Swarm Optimization}

Ant Colony Optimization (ACO) is a robust bio-inspired optimization algorithm based on the foraging patterns of ants in nature. It simulates the behavior of ants in search of the most optimal path between the nest and the food source. It employs the population of artificial ants that traverse through a proposed solution space with the help of pheromone trails that guide future search behavior. The algorithm iteratively updates the global and local pheromone levels based on the solution discovered by ants through every iteration until it reaches an optimal and shortest path. As observed in the given pseudocode, the ACO algorithm uses the collective intelligence of ants and the involved parameters of pheromone levels to solve the optimization problem. Pheromone levels ($\tau_{ij}$) are initialized on all possible solutions (partitions) in order to influence the choices of ants. For every iteration, ants devise solutions based on probabilistically choosing partitions. This is done based on pheromone levels and the solution's quality ($\text{fitness}_{ij}$). The probability of an ant choosing partition $(A_i, B_j)$ is calculated using Equation $8$.

\begin{equation}
P_{ij} = \frac{(\tau_{ij})^\alpha \cdot (\text{fitness}_{ij})^\beta}{\sum_k (\tau_{ik})^\alpha \cdot (\text{fitness}_{ik})^\beta}
\end{equation}

Here, $\alpha$ and $\beta$ control the role and importance of fitness and pheromone in decision-making. 

The step of pheromone updating the chosen partition after the ant tour is essential to achieve the algorithm's convergence. The pheromone update equation is given in Equation $9$. 

\begin{equation}
\tau_{ij} = (1 - \rho) \cdot \tau_{ij} + \sum_{\text{ant}} \Delta \tau_{ij}^\text{ant}
\end{equation}

where $\Delta \tau_{ij}^\text{ant}$ represents the pheromone deposit made by each ant and $\rho$ is the pheromone evaporation rate.
The summarized pseudocode of the Ant Colony Swarm Optimization Algorithm is displayed in Algorithm $6$. 

\begin{algorithm}
\caption{Ant Colony Optimization}
\SetKwInOut{Input}{Input}
\SetKwInOut{Output}{Output}
\Input{Number of Ants $N$, Number of Iterations $\text{MaxIter}$,
       Pheromone Evaporation Rate $\rho$, Initial Pheromone $\tau_0$,
       Ant Movement Control Parameters $\alpha$ and $\beta$}
\Output{Best Solution}
Initialize pheromone levels $\tau_{ij}$ on all edges $(i, j)$ to $\tau_0$\;
Initialize bestSolution with an arbitrary solution\;
Initialize bestObjective with a large value\;
\For{$\text{iter} = 1$ \KwTo $\text{MaxIter}$}{
    \For{$\text{ant} = 1$ \KwTo $N$}{
        Initialize ant's currentSolution with an arbitrary solution\;
        \For{each step in the solution}{
            Calculate the probability of moving to each neighboring solution based on pheromone\;
            Choose the next solution using the probability distribution\;
            Update ant's currentSolution and currentObjective\;
        }
        \If{currentObjective is better than bestObjective}{
            Update bestSolution and bestObjective\;
        }
        Update pheromone levels on all edges based on ant's tour and evaporation rate $\rho$\;
    }
}
Return bestSolution\;
\end{algorithm}
Through exploration and exploitation, ACO searches the solution space reasonably well to generate solutions while also updating the pheromone trails to converge toward a near-optimal solution. It emulates the process of slowly cooling material to achieve a low-energy ground state. It, therefore, tries to explore the solution spaces by gradually decreasing the temperature parameter. Thus the primary objective of the SA optimization algorithm is to escape the local optima, exploring the solution space extensively.

\section{Proposed Approach}
This section describes the implementation of metaheuristic algorithms, QAOA, and the annealing implementations from Section 2. As mentioned above, we will be using the number partitioning problems to apply the optimizations in order to verify our approach.  The number partitioning problems were randomly generated using the OpenQAOA SDK \cite{sharma2022openqaoa}. Although OpenQAOA also provides QAOA circuit creation features and capabilities, these were ignored in order to implement the custom optimizers. 

Using the random problems created using the SDK, number partitioning QUBOs were created. Given an array $A$ of $n$ elements $[a_{1}, a_{2}, a_{3}, ... a_{n}]$, the Number Partitioning QUBO would be be:
\begin{equation}
    Q = (c-2\sum_{i=1}^{n}a_{i}x_{i})^2
\end{equation}
Where $c$ is the sum of the elements of the array and $x_{i}$ is a binary quadratic variable that could take the values $0$ or $1$ depending on which set $a_{i}$ gets partitioned into.

Once the QUBO problems are formulated, the next steps depend on which algorithm is used:

\subsection{Quantum Annealing Implementation}
Once the QUBO model was created using the DWAVE API \cite{boothby2020nextgeneration}, it was converted to the form of a binary quadratic model and then run on the Quantum Annealing Sampler. From that, the first $1000$ samples were taken.

The solution with the lowest energy is then taken, converted from binary variables to a partition of values, and then the quality of the solution is measured. The quality metric $q$ for a number partitioning solution into two sets $S_1$ and $S_2$ with sums $b_1$ and $b_2$ respectively is:
\begin{equation}
    R =\frac{\max(b_1, b_2)}{c/2}
\end{equation}

Thus the optimal partition should have $R=1$ and near-optimal partitions would minimize $q-1$. The time was measured using the Quantum Processing Unit (QPU) access time accessible from the DWAVE Solver. QPU plays a fundamental part in order to measure the performance of the optimization. 

\subsection{QAOA Implementations}
The QAOA circuit was constructed using instructions provided by the IBM Qiskit Textbook\cite{Qiskit}.\newline 

\noindent When rewriting the earlier mentioned QUBO in the form: 
\begin{equation}
    \sum_{i, j=1}^nx_{i}Q_{ij}x_{j} + \sum_{i=1}^{n}c_{i}x_{i}
\end{equation}
the cost hamiltonian was constructed by applying a $RZ$ gate across all qubits with angle $\theta_{1}$ according to equation:
\begin{equation}
    \theta_{1} = \frac{1}{2}(\sum_{i=1}^{n}c_{i}+\sum_{i=1}^{n}Q_{ij})\gamma
\end{equation}
and a $RZZ$ gates across all pairs of qubits with angle $\theta_{2}$ according to:
\begin{equation}
    \frac{1}{4}Q_{ij}\gamma
\end{equation}
Similarly, the mixer hamiltonian was constructed by applying $RX$ gates across all qubits with angle $\theta_{3}$ with:
\begin{equation}
    \theta_{3} =  2\beta
\end{equation}
Note that all $\gamma$ and $\beta$ are the circuit parameters optimized during the optimization phase. All circuits had $2$ layers of cost and mixer Hamiltonians and were initialized to random ansatz with the $\gamma$ and $\beta$ parameters chosen from $(-\pi, \pi)$.
After the circuit construction, the optimization sequence was the implementations for the GA, DE, PSO, and ACO metaheuristics described in the background section. Using the Number Partitioning Objective Function described under annealing bitstrings were mapped to energies and the optimizers minimized these energies. After optimization was complete and the time was measured $s$ samples were taken where $s$ depends on the equation:
\begin{equation}
    s = \lfloor \log_{10} 2^n \rfloor
\end{equation}
The final solution was the best solution using the Number Partitioning Metric. The optimizer used for the standard QAOA circuit was the Constrained Optimization BY Linear Approximations (COBYLA) which is commonly cited as one the most commonly used and best optimizers \cite{FERNANDEZPENDAS2022113388} and the remaining custom optimizers were implemented with $10$ individuals over $50$ iterations according to the details in the background.

\section{Results and Discussion}

All measures for times were taken in seconds. The times and accuracy were computed as an average of $5$ randomly sampled inputs across 4, 8, and 12 input variables. All tests were done on the QASM Simulator using default settings. Future research will investigate the impact of noise. As shown in the figure below, the metaheuristic optimizers did not improve upon QAOA in terms of time complexity. Quantum Annealing and standard QAOA were by far the fastest algorithms followed by QAOA using Genetic Algorithm and Particle Swarm Optimization which had almost the same time performance, then QAOA using Differential Evolution, and finally, QAOA using Ant Colony Swarm Optimization. The time taken by various metaheuristic optimizers to random NPP instances is plotted in figure $1$ below.\newline

\begin{figure}[h]
    \centering
    \includegraphics*[width=\textwidth]{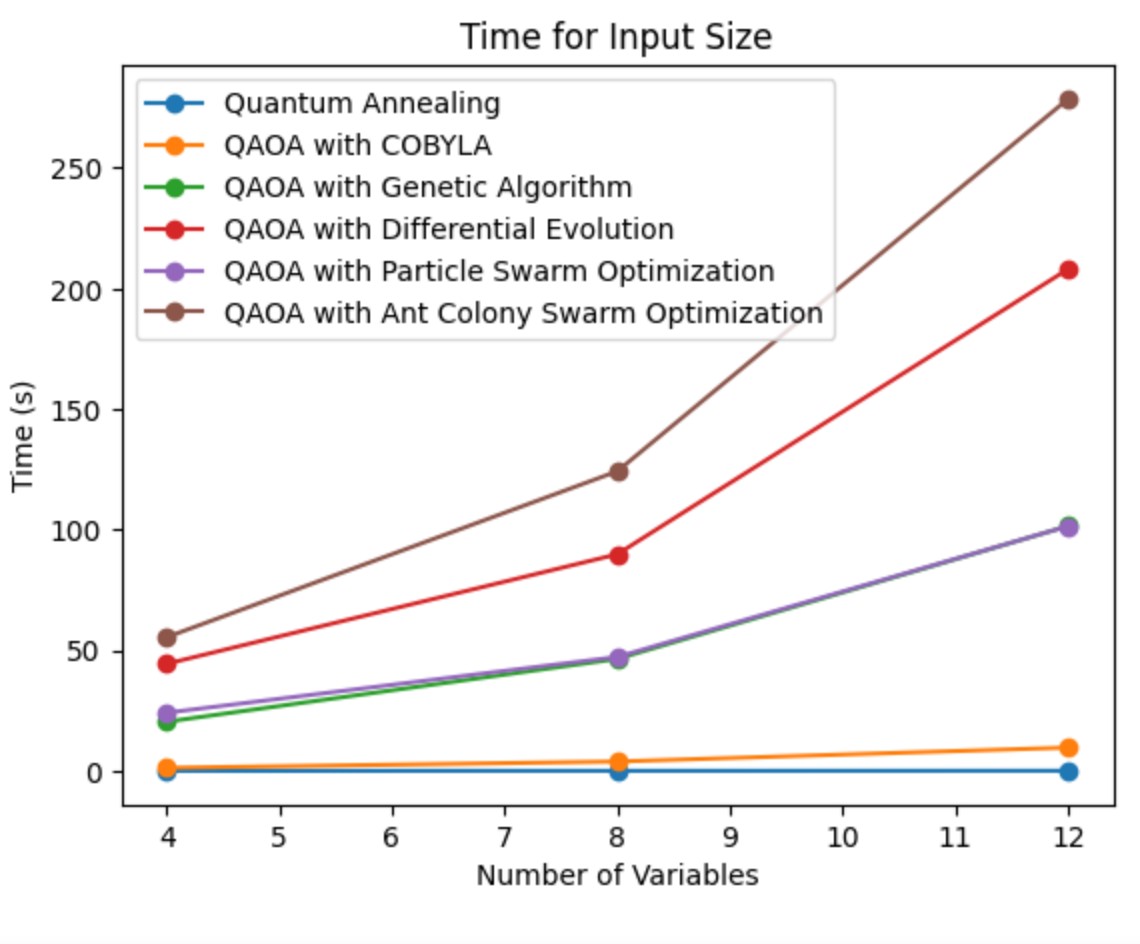}
    \caption{Times in seconds for various metaheuristic optimizers applied to random NPP instances with $4$, $8$, $12$ variables}
    \label{fig:enter-label}
\end{figure}
This research suggests that metaheuristic optimizers can significantly enhance the performance of quantum circuits like QAOA, particularly in terms of accuracy. All metaheuristic optimizers performed significantly better in terms of accuracy compared to standard QAOA circuits. As shown below in Figure 2, QAOA using Ant Colony Swarm Optimization had the best accuracy almost matching quantum annealing, followed by Differential Evolution, then Particle Swarm and Genetic Algorithm, and finally standard QAOA. This suggests that ACSO is effective at finding high-quality solutions. Overall, all metaheuristic-enhanced QAOA circuits outperformed the standard QAOA circuits in terms of accuracy, so further optimized implementations for time complexity may be valuable in the future. While quantum annealing remains the best quantum algorithm for optimization problems, the results suggest that tailored metaheuristic-enhanced quantum algorithms can be competitive in terms of accuracy. The accuracy for various input sizes is plotted with respect to $R-1$ Approximation Ratios in figure $2$ shown below.
\begin{figure}[htbp]
    \centering
    \includegraphics*[width=\textwidth]{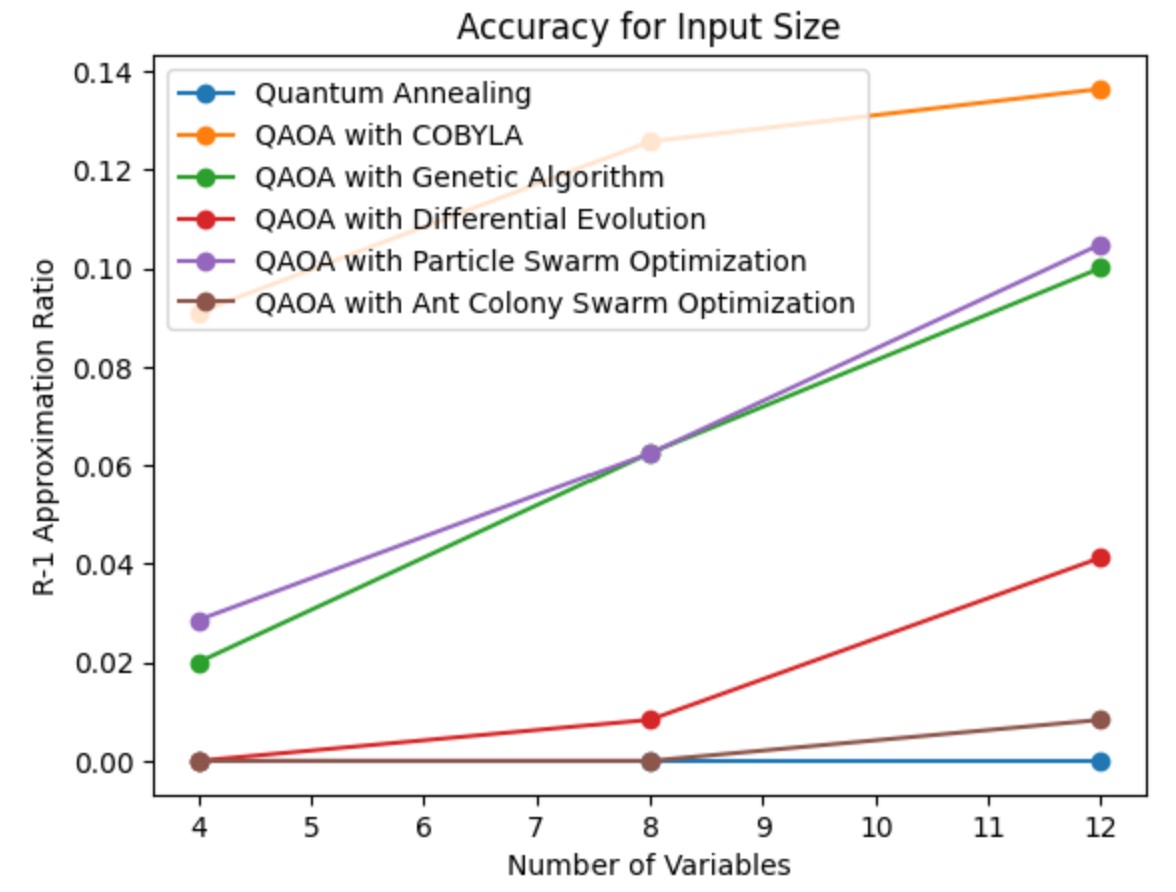}
    \caption{$R-1$ Approximation Ratios for various metaheuristic optimizers applied on QAOA circuit. Lower values indicate more accurate and robust optimizers.}
    \label{fig:sunset}
\end{figure}

When plotting the accuracy of optimizers, the NPP approximation metric computed above was then subtracted by $1$. This was to show the relative percent difference between the obtained solution and the optimal solution. Lower values indicate more accurate optimizers. Metaheuristic algorithms (ACSO, DE, GA, PSO) consistently outperform standard QAOA. The results clearly indicate that ACSO is the most suitable algorithm among all the other algorithms investigated. The plausible explanation is that ACSO can be robust in solving problems related to combinatorial optimization due to its ability to explore the search space and converge to optimum solutions over time. However, it is observed that ACSO's performance can be sensitive to its parameter settings such as that of pheromone update rules \& and exploration-exploitation balance. Tuning these parameters proved to be a challenge while testing. While DE proved to be versatile due to its simplicity and effectiveness. DE may not have performed as well as ACSO in problems with complex discrete search spaces such as NPP, however, DE still proved to handle discrete problems efficiently with appropriate modifications.
PSO may perform better than GA due to its ability to efficiently explore the search space. PSO's inherent exploration mechanism helps it navigate the search space effectively and avoid getting stuck in local minima, unlike GA. Standard QAOA's lower accuracy can be possibly attributed to inherent challenges such as noise, limited qubit resources, and the variational nature of the algorithm.

\section{Conclusion}

In this paper, we design and implement four novel metaheuristic-optimized QAOA circuits. We attempt to incorporate the quantum speedup from parallelism with the exploitation-exploration accuracy benefits of metaheuristic algorithms. Our research provides a novel insight into the applications of metaheuristic-boosted QAOA circuits and their comparisons against Quantum Annealing. Although quantum speedup benefits were lost, and the metaheuristic circuit timing is significantly worse than standard QAOA and not comparable to Quantum Annealing, the accuracy was dramatically improved using the metaheuristics. As shown Differential Evolution and Ant Colony Swarm Optimization almost matched Quantum Annealing in terms of accuracy. Future research can seek to improve upon the metaheuristics through the usage of adaptive algorithms which change their parameters depending on the problem landscape \cite{K_bler_2020}. Further testing will be done on noisy quantum machines to test if performance tapers with quantum noise. Finally experimenting with different $p$ values, iterating through different numbers of cost and mixer Hamiltonians would be useful in trading off the accuracy and time advantages of these metaheuristic optimizers. Overall this paper serves as a valuable reference for classical optimization techniques to improve current NISQ-Era algorithms.

%

\end{document}